\def\half{\frac{1}{2}}
\def\({\left (}
\def\){\right )}
\def\]{\right]}
\def\<{\left <}
\def\>{\right>}
\def\qed{\par\noindent\rightline{$\square$}}
\documentclass[a4paper,12pt]{article}
\usepackage{amssymb, amsmath}
\makeatletter
\renewcommand{\section}{{\setcounter{equation}{0}}\@startsection%
{section}%
{1}%
{0mm}%
{-\baselineskip}%
{0.5\baselineskip}%
{\normalfont\large\bfseries}%
} \makeatother

\makeatletter
\renewcommand{\subsection}{\@startsection%
{subsection}%
{2}%
{0mm}%
{-\baselineskip}%
{0.5\baselineskip}%
{\normalfont\normalsize\bfseries}}%
\newcommand{\ds}{\displaystyle}
\newcommand{\ben}{\begin{enumerate}}
\newcommand{\een}{\end{enumerate}}
\newcommand{\be}{\begin{equation}}
\newcommand{\ee}{\end{equation}}
\newcommand{\bea}{\begin{eqnarray}}
\newcommand{\eea}{\end{eqnarray}}
\newcommand{\beas}{\begin{eqnarray*}}
\newcommand{\eeas}{\end{eqnarray*}}
\newcommand{\begth}{\begin{theorem}}
\newcommand{\enth}{\end{theorem}}
\newcommand{\blem}{\begin{lemma}}
\newcommand{\elem}{\end{lemma}}
\newcommand{\non}{\nonumber}
\newcommand{\nl}{\newline}

\newtheorem{proposition}{Proposition}
\newtheorem{theorem}{Theorem}[section]
\newtheorem{lemma}{Lemma}[section]

\newenvironment{proof}{{\bf Proof:}}{\hfill$\square$\vskip.5cm}
\def\RR{\mathbb{R}}
\def\NN{\mathbb{N}}

\def\PP{\mathbb{P}}

\def\ZZ{\mathbb{Z}}

\def\KK{\mathbb{K}}

\def\tr{{\rm trace}\ }
\def\la{\langle}
\def\ra{\rangle}
\def\veps{\varepsilon}
\def\sQ{{\hbox {\tiny {\sl Q}}}}
\def\sLambda{{\hbox {\tiny{$\Lambda$}}}}
\def\bq{{\bf q}}
\def\sk{\hskip 0.07cm}

\begin{document}
\phantom{.} \textbf{24-03-05}:
\vskip1.5cm
\begin{center}
{\Large Long Cycles in a Perturbed Mean Field Model of a Boson Gas}
\vskip 0.5cm
{\bf Teunis C. Dorlas}
\linebreak
School of Theoretical Physics
\linebreak
Dublin Institute for Advanced Studies
\linebreak
10, Burlington Road, Dublin 4, Ireland
\linebreak
email:dorlas@stp.dias.ie
\vskip 0.5cm
{\bf Philippe A. Martin}
\linebreak
Institut of Theoretical Physics
\linebreak
Swiss Federal Institute for Technology (EPFL)
\linebreak
CH 1015 Lausanne, Switzerland
\linebreak
email: Philippe-Andre.Martin@epfl.ch
\vskip 0.5cm
and
\vskip 0.5cm
{\bf
Joseph V. Pul\'e}\footnote{{\it
Research Associate, School of Theoretical Physics, Dublin
Institute for Advanced Studies.}}
\linebreak
Department of Mathematical Physics
\linebreak
University College Dublin\\Belfield, Dublin 4, Ireland
\linebreak
email:
Joe.Pule@ucd.ie
\end{center}
\vskip 1cm
\begin{abstract}
\vskip -0.4truecm
\noindent
In this paper we give a precise mathematical formulation of the relation
between Bose condensation and long cycles and prove its validity for  the perturbed mean field model
of a Bose gas.
We decompose the total density $\rho=\rho_{{\rm short}}+\rho_{{\rm long}}$
into the number density of particles belonging to cycles of finite length ($\rho_{{\rm short}}$)
and to infinitely long cycles ($\rho_{{\rm long}}$) in the thermodynamic limit. For this model we
prove that when there is Bose condensation, $\rho_{{\rm long}}$ is different from zero and identical
to the condensate density. This is achieved through an application of the theory of
large deviations. We discuss the possible equivalence of
$\rho_{{\rm long}}\neq 0$ with off-diagonal long range order and winding paths
that occur in the path integral representation of the Bose gas.
\end{abstract}
\newpage\setcounter{page}{1}
\section{Introduction}
In 1953, Feynman analyzed the partition function of an interacting Bose gas in terms of the statistical distribution
of permutation cycles of particles and emphasized the roles of long cycles at the
transition point \cite{Fey 1953 and stat.mech.}. Then Penrose and Onsager, pursuing Feynman's arguments,
observe that there should be Bose condensation when the fraction of the total particle number belonging to
long cycles is strictly positive \cite{Pen.Onsa.}.
These ideas are now commonly accepted and also discussed in various contexts in systems showing analogous phase
transitions
such as percolation, gelation and polymerization (see e.g. \cite{Chandler}, \cite{Sear-Cuesta}, \cite{Schakel}).
However, to our knowledge, there had not appeared a precise mathematical and quantitative formulation of the relation
between Bose condensate and long cycles until  the the recent work of S\"ut\"o \cite{Suto} dealing with the
free and mean field gas.
The purpose of this paper is to formulate such a relation and to prove its validity in a model of a Bose gas having
some genuine short range interactions between modes, the perturbed mean field model studied in \cite{vdB D L P}.
\par
In Section 2 we recall the general framework for cycle statistics independently of any specific model.
The arguments used below have been well known for a long time in various contexts,
see e.g. Ginibre in \cite{Ginibre} for a proof of the convergence of virial expansions of quantum gases,
Cornu \cite{Cornu} in relation to Mayer expansions for quantum Coulomb systems, Ceperley \cite{Ceperley}
for numerical simulations on Helium; a pedagogical account can be found in \cite{Martin}.
Using standard properties of the decomposition of permutations into cycles, the grand-canonical sum is
converted into a sum on cycle lengths.
This makes it possible to decompose the total density $\rho=\rho_{{\rm short}}+\rho_{{\rm long}}$
into the number density of particles belonging to cycles of finite length ($\rho_{{\rm short}}$)
and to infinitely long cycles ($\rho_{{\rm long}}$) in the thermodynamic limit. It is conjectured
that when there is Bose condensation, $\rho_{{\rm long}}$ is different from zero and identical
to the condensate density. The main result of the paper is to establish the validity of this conjecture
in the perturbed mean field model.
\par
This model is diagonal with respect to the mode occupation numbers with a non mean field
interaction between them. It has been thoroughly analyzed in \cite{DLP 1}, \cite{DLP 2} and it is revisited here
in Sections 3 to 6 from the view point of cycle statistics. In order to compute
$\rho_{{\rm long}}=\rho-\rho_{{\rm short}}$, we find it useful to introduce a modified statistical ensemble
equipped with an additional external field that gives an extra weight to cycles of finite length.
Then  $\rho_{{\rm short}}$ is related to the derivative of the pressure with respect to this external field.
In Section 3 the mean perturbed field model is presented, and an easy explicit calculation
using the modified ensemble establishes the identification of $\rho_{{\rm long}}$
with the  Bose condensate for the free gas. Section 4 is devoted to the formulation
of the variational principle that leads to the same identification for the interacting gas.
The justification of this variational principle is provided by an application of the theory of
large deviations (Section 5) and the mathematical proofs are found in Section 6.
In the concluding remarks (Section 7) we discuss the possible equivalence of
$\rho_{{\rm long}}\neq 0$ with off-diagonal long range order and winding paths
that occur in the path integral representation of the Bose gas.

\section{Cycle statistics}

\subsection{Cycle representation of the partition function}

We consider a system of $n$ identical bosons enclosed in a region $\Lambda$ in thermal equilibrium at
temperature $k_{B}T=\beta^{-1}$. The states of these bosons belongs to the symmetrization
${\cal H}^n_{\sLambda,\ {\tiny \rm symm}}$ of the $n$-fold tensor product of
${\cal H}^{\otimes n}_\sLambda$ of the one particle Hilbert space
${\cal H}_\sLambda$ and their total energy is given by a symmetric Hamiltonian $H_{\Lambda}$.
The projection   $P^n_{{\tiny \rm symm}}$ of ${\cal H}^{\otimes n}_{\sLambda}$ onto
${\cal H}^n_{\sLambda,\ {\tiny \rm symm}}$ reads
\be
P^n_{{\tiny \rm symm}}=\frac{1}{n!}\sum_{\pi\in S_n}U_\pi.
\label{proj}
\ee
where  $U_\pi:{\cal H}^{\otimes n}_{\sLambda}\mapsto{\cal H}^{\otimes n}_{\sLambda}$ is a unitary
representation  of the permutation group  $S_n$ on ${\cal H}^{\otimes n}_{\sLambda}$ defined by
\be
U_\pi(\phi_1\otimes\phi_2\otimes \dots \otimes \phi_n)
=\phi_{\pi(1)}\otimes\phi_{\pi(2)}\otimes \dots \otimes \phi_{\pi(n)},
\quad \phi_j \in  {\cal H}_\sLambda,\quad
j=1,\ldots,n,\;\pi\in S_{n}.
\ee
The canonical partition function can be written as
\bea
Z _{\sLambda}^{n}=\tr_{{\cal H}^n_{\sLambda,{\tiny \rm symm}}}\left[e^{-\beta H_{\Lambda}}\right]&
=&\tr_{{\cal H}^{\otimes n}_\sLambda}
\left[P^n_{{\tiny \rm symm}}e^{-\beta H_{\sLambda}}\right]=\frac{1}{n!}\sum_{\pi\in S_n}f_{\sLambda}(\pi)
\label{0.1}
\eea
where $f_{\sLambda}(\pi)$ is a function on $S_{n}$ defined by
\be
f_{\sLambda}(\pi)=\tr_{{\cal H}^{\otimes n}_\sLambda}\left[U_{\pi}
e^{-\beta H_{\Lambda}}\right].
\label{0.2}
\ee
\par
Each permutation $\pi \in S_n$ can be decomposed into a number of cyclic permutations of lengths
$q_1,q_2,\dots,q_r$ with $r\leq n$ and  $q_1+q_2+\dots+q_r=n$. We consider the set
${\ds \Omega=\cup_{r\in\NN}\Omega_r}$ of unordered $r$-tuples of natural numbers
${\bf q}=[q_1,q_2,\dots,q_r]\in \Omega_r$ for $r=0,1,2,\ldots$, and let
$|{\bf q}|=q_1+q_2+\dots+q_r$ for ${\bf q}\in  \Omega$.
Then a decomposition  of $\pi\in S_n$ into cycles is labelled by  ${\bf q}\in  \Omega$ with $|{\bf q}|=n$.
We recall the following facts on the permutation group.
\begin{itemize}
\item The decomposition into cycles leads to a partition of $S_n$
into equivalence classes of permutations $C_{{\bf q}},\;|{\bf q}|=n$.
\item  Two permutations $\pi'$ and $\pi''$ belong to the same class
if and only if they are conjugate in $S_n$, i.e. if there exists $\pi \in S_n$ such that
\be
\pi''=\pi^{-1}\pi '\pi \;.
\label{0.3}
\ee
\item The number of permutations belonging to the class $C_{{\bf q}}$ is
\be
\frac{n!}{n_{{\bf q}}!(q_1q_2\dots q_r)}\
\label{0.4}
\ee
with $n_{{\bf q}}!=n_{1}! n_{2}!\cdots n_{j}! \cdots$
and $ n_{j}$ is the number of cycles of length $j$ in  ${\bf q}$.
\end{itemize}
\par
We observe that for a symmetric Hamiltonian ($[H_{\sLambda}, U_{\pi}]=0,\,\pi\in S_n$), $f_{\sLambda}(\pi)$ is
constant on
$C_{{\bf q}}$. Indeed for $\pi',\pi''\in C_{{\bf q}}$, one has by (\ref{0.3})
\bea
f_{\sLambda}(\pi'')&=&\tr_{{\cal H}^{\otimes n}_\sLambda}\left[U_{\pi''}
e^{-\beta H_{\Lambda}}\right]
=\tr_{{\cal H}^{\otimes n}_\sLambda}\left[U_{\pi}^{-1}U_{\pi'}U_{\pi}e^{-\beta H_{\Lambda}}\right]
\nonumber\\ &=&\tr_{{\cal H}^{\otimes n}_\sLambda}\left[U_{\pi}^{-1}U_{\pi'}e^{-\beta H_{\Lambda}}U_{\pi}\right]=
\tr_{{\cal H}^{\otimes n}_\sLambda}\left[U_{\pi'}
e^{-\beta H_{\Lambda}}\right]=f_{\sLambda}(\pi') \;.
\label{0.5}
\eea
Therefore $f_{\sLambda}(\pi)$ only depends
on the cycle decomposition $\bq$ of $\pi$, and we shall write from now on
\be
f_{\sLambda}(\pi)=f_{\sLambda}(\bq),  \,\; \pi\in  C_{{\bf q}}.
\label{0.5a}
\ee
For instance $f_{\sLambda}(\bq)=f_{\sLambda}(\pi_\bq)$ can be evaluated on the  permutation
$\pi_{{\bf q}}=(1,\ldots, q_1)(q_1+1,\ldots, q_1+q_2)\ldots(n-q_r,\ldots, n)$ representative of $C_{{\bf q}}$.
Let $U_{q}$ be the operator representing the cyclic  permutation $(1,2,\ldots ,q)$. Then
\be
U_{\pi_{{\bf q}}}=U_{q_{1}}U_{q_{2}}\cdots U_{q_{r}},
\label{0.5b}
\ee
\be
f_{\sLambda}(\bq)=\tr_{{\cal H}^{\otimes n}_\sLambda}\left[U_{q_{1}}U_{q_{2}}\cdots U_{q_{r}}e^{-\beta H_{\Lambda}}\right]
\ee
 and using (\ref{0.4})
\be
\sum_{\pi\in {\rm C}_{{\bf q}}}f_{\sLambda}(\pi)=\frac{n!}{n_{{\bf q}}!(q_1q_2\dots q_r)}f_{\sLambda}(\bq)\;.
\label{0.6}
\ee
Then, summing first in (\ref{0.1}) on the permutations belonging to a class $C_{{\bf q}}$ and then on all
possible classes gives
\bea
Z_{\sLambda}^{n}&=&\frac{1}{n!}\sum_{r=1}^n\ \sum_{{\bf q}\in \Omega_r,\, |{\bf q}|=n}
\ \sum_{\pi\in {\rm C}_{{\bf q}}}f_{\sLambda}({\bf q})\non\\
&=&\frac{1}{n!}\sum_{r=1}^n\frac{1}{r!}\ \sum_{{\bf q}\in \NN^r,\, |{\bf q}|=n}n_{{\bf q}}!\
\sum_{\pi\in {\rm C}_{{\bf q}}}f_{\sLambda}({\bf q})\non\\
&=&\sum_{r=1}^n\frac{1}{r!}\ \sum_{{\bf q}\in \NN^r,\,
|{\bf q}|=n}\frac{1}{(q_1q_2\dots q_r)}f_{\sLambda}(q_1,q_2,\dots,q_r)
\label{0.7}
\eea
In the second line we have taken into account that replacing the summation on unordered
$r$-tuples in $\Omega_r$ by ordered $r$-tuples in  $\NN^r$,
we are over counting by a factor $r!/n_{{\bf q}}!$. Finally to get rid of the constraint of fixed particle number $n$,
we introduce the grand-canonical partition function  with chemical potential $\mu$
\be
\Xi^\mu_\sLambda=1+\sum_{n=1}^{\infty}e^{\beta\mu n}Z_{\sLambda}^{n}
=1+\sum_{r=1}^{\infty}\frac{1}{r!}\ \sum_{{\mathbf q}\in\NN^r}\
\prod_{i=1}^r\frac{e^{\beta\mu q_i}}{q_i} f_\sLambda(q_1,q_2,\dots,q_r)\;.
\label{0.8}
\ee
This leads to define the grand-canonical distribution of cycles on the space
$\Omega=\cup_{r\in\NN}\Omega_{r}$ of $r$-tuples of arbitrary integer length by the  probability measure
\bea
\PP^\mu_\sLambda({\bf q})=\frac{1}{\Xi^\mu_\sLambda}\frac{1}{r!}\prod_{i=1}^r\frac{e^{\beta\mu q_i}}{q_i}
f_\sLambda(q_1,q_2,\dots,q_r),\quad \bq \in \Omega, \label{0.9}\\
\sum_{r=1}^{\infty}\sum_{q_1,q_2,\dots,q_r \in \NN^r}\PP^\mu_\sLambda (q_1,q_2,\dots,q_r)=1\nonumber
\eea
One then introduces the cycle density and cycle correlations in the usual way.
Consider the observable $\ds{\frac{1}{|\Lambda|}\sum_{i=1}^{r}\delta_{q,q_{i}}}$ i.e.
the number of cycles of length $q$ per unit volume. Its grand-canonical average with respect to
$\PP^\mu_\sLambda$ is given by
\bea
\rho^\mu_\sLambda(q)&=&\<\frac{1}{|\Lambda|}\sum_{i=1}^{r}\delta_{q,q_{i}}\>_{\PP^\mu_\sLambda}\nonumber\\
&=&\frac{1}{|\Lambda|}\ \frac{1}{\Xi^\mu_\sLambda}\ \frac{e^{\beta\mu q}}{q}\ \sum_{r\in\NN}\frac{1}{(r-1)!}\
\sum_{{\mathbf q}\in\NN^{(r-1)}}\ \prod_{i=1}^{r-1}\frac{e^{\beta\mu q_i}}{q_i} f_\sLambda(q,q_1,\dots,q_{(r-1)})\;.
\label{0.10}
\eea
The quantities
\be
P_\sLambda(\mu)=\frac{1}{\beta |\Lambda|}\ln\Xi^\mu_\sLambda\quad {\rm and}\quad
\rho^\mu_\sLambda=\sum_{q=1}^{\infty}q\rho^\mu_\sLambda(q)
=\frac{1}{\beta|\Lambda|}\frac{\partial}{\partial \mu}\ln\Xi^\mu_\sLambda
\label{0.11}
\ee
are the usual grand-canonical pressure and particle number density of the gas.
\subsection{Particles in long and short cycles}
We first note that expressing the partition function in terms of cycles corresponds
to a partitioning of the particles into groups of the same size as the cycle lengths.
It is natural therefore to ask for the probability $p_\sLambda(q)$ for a particle to belong to a cycle of length
$q$ (we drop the chemical potential $\mu $ from the notation and
it is understood that $\mu $ and the inverse temperature $\beta$ are kept fixed). Since the average number of
particles to be found in $q$-cycles is $|\Lambda| q\rho_\sLambda(q) $ and the total average particle number
is $|\Lambda| \rho_\sLambda$
this probability is
\be
p_\sLambda(q)=\frac{q\rho_\sLambda(q)}{ \rho_\sLambda}, \quad \quad
\sum_{q=1}^{\infty}p_{\sLambda}(q)=1.
\label{0.12}
\ee
Let us assume at this point that infinite volume limits of the cycle and total densities exist, namely
\be
\lim_{\Lambda\to\infty}\rho_\sLambda(q)=\rho(q),\quad \lim_{\Lambda\to\infty}\rho_\sLambda=\rho\;.
\label{0.13}
\ee
Then $\lim_{\Lambda\to\infty}p_\sLambda(q)=q\rho(q)/\rho=p(q)$ is the probability to find a particle in
a $q$-cycle, $q<\infty$, in the infinitely extended system. Hence the fraction of particles
$\rho_{{\rm long}}/\rho$ belonging to infinitely long cycles is
\be
\frac{\rho_{{\rm long}}}{\rho}=1-\sum_{q=1}^{\infty}p(q)
\label{0.14}
\ee
The conjecture \cite{Pen.Onsa.} is that $\rho_{{\rm long}}$ is different from zero when there is Bose
condensation and identical to the condensate density. One also says
in this case that the probability $p(q)$ is defective or that cycle percolation occurs.
\par
From the above definitions one can write
\bea
\rho_{{\rm long}}&=&\rho-\sum_{q=1}^{\infty}q\rho(q)=\lim_{\Lambda\to\infty}\sum_{q=1}^{\infty}q\rho_\sLambda(q)-
\sum_{q=1}^{\infty}q\rho(q)\non\\
&=&\lim_{\Lambda\to\infty}\sum_{q=Q+1}^{\infty}q\rho_\sLambda(q)-\sum_{q=Q+1}^{\infty}q\rho(q)
\label{0.15}
\eea
for any $Q>0$. Thus letting $Q\to\infty$ and taking into account that the last sum in (\ref{0.15}) is convergent one has
\be
\rho_{{\rm long}}=\lim_{Q\to\infty}\lim_{\Lambda\to\infty}\sum_{q=Q+1}^{\infty}q\rho_\sLambda(q)
\label{0.16}
\ee
showing that $\rho_{{\rm long}}$ is carried by the tail of the cycle number density as $\Lambda\to\infty$.
\par
One can alternatively consider the quantity
\be
\rho_{{\rm short}}=\rho-\rho_{{\rm long}}=\lim_{Q\to\infty}\lim_{\Lambda\to\infty}\sum_{q=1}^Qq\rho_\sLambda(q)
\label{0.17}
\ee
which is expected to coincide with the density of particles that do not belong to the condensate and can be studied more easily.
It will turn out to be convenient to define a modified partition function depending on a parameter $\lambda\geq 0$ by
\be
\Xi_{\sQ,\sk\lambda,\sk\sLambda}=1+\sum_{r\in\NN}\frac{1}{r!}\ \sum_{{\mathbf q}\in\NN^r}\
\prod_{i=1}^r\frac{e^{\beta(\mu + \lambda\theta_Q(q_i))q_i}}{q_i} f_\sLambda(q_1,q_2,\dots,q_r)
\label{0.18}
\ee
where
\be
\theta_\sQ(q)=
\begin{cases}
1 & {\rm if}\ q\leq Q,\\
0 & {\rm if}\ q > Q.\\
\end{cases}
\ee
and the corresponding modified pressure
\be
P_{\sQ,\sk\lambda,\sk\sLambda}=\frac{1}{\beta|\Lambda|}\ \ln\Xi_{\sQ,\sk\lambda,\sk\sLambda}.
\label{0.19}
\ee
One immediately sees that
\be
\sum_{q=1}^Qq\rho_\sLambda(q)=\frac{\partial}{\partial \lambda}P_{\sQ,\sk\lambda,\sk\sLambda}\Big|_{\lambda=0}.
\label{0.20}
\ee
Thus if
$P_{\sQ,\sk \lambda}=\lim_{\Lambda\to \infty}P_{\sQ,\sk\lambda,\sk\sLambda}$
has a thermodynamic limit, the convexity of $P_{\sQ,\sk\lambda,\sk\sLambda}(\mu)$
implies the convergence of the derivative \cite{Griffiths} and
\be
\rho_{{\rm short}}=\lim_{Q\to\infty}\lim_{\Lambda\to\infty}\sum_{q=1}^Qq\rho_\sLambda(q)=\lim_{Q\to\infty}\frac{\partial}{\partial \lambda}P_{\sQ,\sk \lambda}\Big|_{\lambda=0}.
\label{0.22}
\ee
Note that the order of limits is crucial here.
The procedure is the same as generating the density from the pressure by functional differentiation with respect to an external field.
Here the \lq \lq external field" $\lambda\theta_Q(q)$ gives an additional weight to cycles of length $q\leq Q$.

\section{The perturbed mean field model}

The bosons have mass $M$  and are enclosed in a cube $\Lambda\subset \RR^d$ centred at the origin with periodic boundary
conditions. The one-particle space ${\cal H}_\sLambda=L^2(\Lambda)$  is generated by the basis of plane waves
with wave number $k\in \Lambda^*=\{2\pi l/|\Lambda|^{1/d}\ |\  l\in\ZZ^d\}$ where $|\Lambda|$ is the volume of $\Lambda$.
In second quantized form, the Hamiltonian of the perturbed mean field model expressed in terms of the boson creation and annihilation operators
$a^*_k$ and $a_k$, $k\in \Lambda^*$, $[a_k,a^*_{k'}]=\delta_{k,k'}$, reads
\be
H_\Lambda=T_\Lambda+\frac{a}{2V}N^2_\Lambda+\frac{1}{2V}\sum_{k, k'\in \Lambda^*}v(k,k')N_kN_{k'}
\label{0.23}
\ee
where  $N_k=a^*_ka_k$ are the occupation number operators of the modes $k$, and
\be
T_\Lambda=\sum_{k\in \Lambda^*}\epsilon(k)N_k,\quad\epsilon (k)=\hbar^{2}\|k\|^2/2M, \quad N_\Lambda=\sum_{k\in \Lambda^*}{\hskip -0.1cm}N_k
\label{0.24}
\ee
are the kinetic and total particle number operators. It differs from the standard mean field model
by the last term of (\ref{0.23}) which
introduces the coupling kernel $v(k,k')$ between the modes. We assume that $v(k,k')$ is a bounded,
continuous and positive definite kernel
with suitable decay properties.
\par
Since the perturbed mean field model is defined in terms of the mutually commuting occupation numbers
$N_{k}$, it can be studied with the techniques of classical probability theory.
The $N_{k}$'s form a set of classical random variables and the equilibrium properties are expressed
in terms of their distribution functions. In particular, in the
thermodynamic limit, the average occupation number distribution is a measure $\nu(dk)$ which
is continuous except for a possible Dirac $\delta$ at $k=0$ corresponding to a macroscopic
occupation of the $k=0$ state.
Thus when there is a non vanishing condensate density $\nu_{c}$, the equilibrium distribution
is given by
\bea
\nu(dk)=\nu_{c}\delta(k)dk+\nu_{e}(k)\frac{dk}{(2\pi)^{d}}
\label{0.25}
\eea
where the continuous part $\nu_{e}(k)$  describes the density of particles in excited states.
The main result of the paper consists in establishing that the average number density of short
cycles (\ref{0.17})  is identical  with the total density $\nu_{e}=\int \nu_{e}(k)\frac{dk}{(2\pi)^{d}}$
of exited particles, and consequently that $\rho_{{\rm long}}$ (\ref{0.16}) indeed coincides with the
condensate density of particles $\nu_{c}$ in the $k=0$ mode as given in (\ref{0.25}).
\par
For a number of specific interaction kernels it is known that Bose condensation occurs for
a range of chemical potentials. In particular in the case of a Gaussian kernel
$v(k,k')=v_{0}e^{-c||k-k'||^{2}}$ there is a value $\mu_{0}$ such that if $\mu\in (0,\mu_{0})$,
then $\rho_{c}$ is non zero \cite{DLP 1} . On the other hand for the kernel
$v(k,k')=v_{0}e^{-c||k-k'||}$ there is condensation for all positive $\mu$ \cite{DLP 2}.
Our present results therefore imply that in these cases the Bose condensate is carried by particles
in long cycles.{}
\par
We compute the function
$f_\sLambda(q_1,q_2,\dots,q_r)$ for the perturbed mean field model (\ref{0.23}) by  calculating the trace
(\ref{0.2}) in the plane wave basis
\bea
f_{\sLambda}(\pi)&=&\sum_{{\bf k}\in(\Lambda^*)^r }\langle k_{1},\ldots, k_{n}|
e^{-\beta H_\sLambda}|k_{\pi_{(1)}},\ldots,k_{\pi_{(n)}}
\rangle\nonumber\\
&=& \sum _{{\bf k}\in(\Lambda^*)^r}\exp\left\{-\beta\left(\sum_{j=1}^n \veps(k_j) +\frac{a}{2|\Lambda|}n^{2}
+ \frac{1}{2|\Lambda|}\sum_{j,j'=1}^n v(k_j,k_{j'})\right)\right\}\prod_{j=1}^{n}\delta_{k_{j},k_{\pi_{j}}}
\non\\
&&
\label{0.24a}
\eea
giving
\bea
&& f_\sLambda(q_1,q_2,\dots,q_r)=\non\\
&&\hskip 1cm\sum_{{\bf k}\in(\Lambda^*)^r}
\exp\left\{-\beta\left(\sum_{i=1}^r \veps(k_i)q_i + \frac{a}{2|\Lambda|}\(\sum_{i=1}^r q_i\)^2
+ \frac{1}{2|\Lambda |}\sum_{i,i'=1}^r q_iq_{i'}v(k_i,k_{i'})\right)\right\}.\non\\
\label{0.24b}
\eea
As a preliminary simple calculation, we consider the modified partition function
$\Xi^0_{\sQ,\sk\lambda,\sk\sLambda}$ of the free gas. In the absence of interactions the function
 (\ref{0.24b}) reduces to
\be
f_\sLambda(q_1,q_2,\dots,q_r)=\sum_{{\bf k}\in(\Lambda^*)^r}
\exp\left\{-\beta\sum_{i=1}^r \veps(k_i)q_i\right\}.
\ee
Inserting this into
(\ref{0.18}) we obtain the modified partition of the free gas in the form
\be
\Xi^0_{\sQ,\sk\lambda,\sk\sLambda}=1+\sum_{r\in\NN}\frac{1}{r!}\ \sum_{{\mathbf q}\in\NN^r}\ \prod_{i=1}^r\frac{e^{\beta(\alpha + \lambda\theta_Q(q_i))q_i}}{q_i}
\sum_{{\bf k}\in(\Lambda^*)^r}\exp\left\{-\beta\sum_{i=1}^r \veps(k_i)q_i\right\}.
\label{0.26}
\ee
Here the chemical potential, denoted by $\alpha$, has to be negative, and the corresponding pressure is
\be
P^{0}_{\sQ,\sk\lambda,\sk\sLambda}=\frac{1}{\beta|\Lambda|}\ \ln\Xi^0_{\sQ,\sk\lambda,\sk\sLambda}
\label{1.2}
\ee
This partition function is readily calculated, permuting product with the ${\mathbf q}$ and ${\bf k}$
summations one obtains
\bea
\Xi^0_{\sQ,\sk\lambda,\sk\sLambda}&=1+&\sum_{r\in\NN}\frac{1}{r!}\sum_{{\bf k}\in(\Lambda^*)^r}\sum_{{\mathbf q}\in\NN^r}
\prod_{i=1}^r \frac{\exp\left\{\beta(\alpha-\veps(k_i)+\lambda\theta_Q(q_i))q_i\right\}}{q_i}\nonumber\\
&=&
\exp\left\{\sum_{ k\in\Lambda^*}\pi_{\sQ,\sk \lambda}(\alpha -\veps(k))\right\}
\label{0.27}
\eea
where for $ y<0$ and $ \lambda\geq 0$ we have introduced the function
\be
\pi_{\sQ,\sk \lambda}( y)=\sum_{q=1}^\sQ \frac{e^{\beta( y+ \lambda)q}}{\beta q}+\sum_{q=\sQ+1}^\infty \frac{e^{\beta y q}}{\beta q}.
\label{0.28}
\ee
This yields the modified pressure (\ref{0.19}) in the infinite volume limit
\be
P^{0}_{\sQ,\sk \lambda}=\lim_{\Lambda\to\infty}P^{0}_{\sQ,\sk\lambda,\sk\sLambda}=\lim_{\Lambda\to\infty}\frac{1}{|\Lambda|}\sum_{k\in \Lambda^*}\pi_{\sQ,\sk \lambda}(\alpha -\veps(k))=\int\frac{dk}{(2\pi)^{d}}\pi_{\sQ,\sk \lambda}(\alpha -\veps(k))\;.
\label{0.29}
\ee
Notice that for $\lambda=0$,
\be
\pi_{\sQ,\sk \lambda=0}( y)\equiv\pi_{0}(y)=\beta^{-1}\ln (1-e^{-\beta y}),
\label{0.30}
\ee
so that
\be
P^{0}_{\sQ,\sk \lambda=0}= \beta^{-1}\int \frac{dk}{(2\pi)^{d}}\ln \(1-e^{-\beta(\veps(k)-\alpha)}\)
\ee
reduces to the well known pressure of the free gas at chemical potential $\alpha < 0$.
Moreover, one explicitly checks from (\ref{0.28}) and  (\ref{0.29}) that
\be
\lim_{Q\to\infty}\frac{\partial}{\partial \lambda}P^{0}_{\sQ,\sk \lambda}\Big|_{\lambda=0}=\int \frac{dk}{(2\pi)^{d}}\frac{1}{e^{\beta(\epsilon(k)-\alpha)}-1}=\nu^{0}_{e}
\label{0.31}
\ee
with $\nu^{0}_{e}$ the total number of particles in exited states. Bose condensation is characterized in the three dimensional gas by the fact that $\nu^{0}_{e}$
remains finite for $\alpha=0$ (the critical density). The relation (\ref{0.31}) also remains valid for $\alpha=0$ showing that the equality $\rho_{{\rm short}}=\nu^{0}_{e}$ holds even in case  when the condensate density is different from zero .
This yields again the results of \cite{Suto} for the free gas.

\section{The variational principle}

We set up a variational principle to determine the modified pressure (\ref{0.19})
in presence of the interaction. Its precise mathematical justification is provided by the application
of the theory of large deviations presented in the next sections.
Here  we give the form of the grand-potentials per unit volume
$\varphi=u-\mu \rho -Ts=-P$ ($u$ is the internal energy, $\rho$
the particle number density and $s$ the entropy) corresponding to the perturbed mean field model and to the modified statistical ensemble (\ref{0.18}). These potentials will be considered
as a functionals of  all possible momentum distributions.
The latter are described by positive bounded measures $m(dp)$ on $\RR^d$.
Their minimization will yield the corresponding pressures.
\par
Consider first the measures of the form
\bea
m_{({\bf q},{\bf k})}(dp)=\frac{1}{|\Lambda|}\sum_{i=1}^{r}q_{i}\delta(k_{i}-p)dp,
\quad ({\bf q},{\bf k})=\{(q_1,k_1),(q_2,k_2)\dots,(q_r,k_r)\}
\label{1.1}
\eea
which are the sum of Dirac distributions centered at the wave numbers $k_{i}$. The particle number density is $\int m_{({\bf q},{\bf k})}(dp)=\sum_{i=1}^{r}q_{i}/|\Lambda|=n/|\Lambda|$, and
$\int_{\Delta}m_{({\bf q},{\bf k})}(dp)$ is the number of particles per unit volume that have wave numbers in the range $\Delta$.
One can write (\ref{0.24b}) as
\bea
f_\sLambda({\bf q})&=&\sum_{{\bf k}\in(\Lambda^*)^r}\exp(-\beta |\Lambda| u_{({\bf q},{\bf k})})\nonumber\\
u_{({\bf q},{\bf k})}&=&\la\epsilon,m_{({\bf q},{\bf k})}\ra+\frac{a}{2}\la1,m_{({\bf q},{\bf k})}\ra^{2}
+\frac{1}{2}\la m_{({\bf q},{\bf k})},vm_{({\bf q},{\bf k})}\ra
\label{1.2b}
\eea
with the notation $\la g,m \ra=\int  g(p)m(dp)$, $\la m,vm\ra=\int v(p,p')m(dp)m(dp')$, and where
$u_{({\bf q},{\bf k})}$ is the energy density of particles in a cycle and wave number configuration $({\bf q},{\bf k})$.
In view of (\ref{1.2b}), we define the energy density functional for general measures $m(dp)$   by
\be
u(m)=\la\epsilon,m \ra +\frac{a}{2}\la 1,m\ra^{2}+\frac{1}{2}\la m,vm\ra\; .
\label{1.3}
\ee
Concerning the entropy density functional, we recall its standard form \cite{Landau}
\bea
s_{0,\Lambda}=-k_{B}\frac{1}{ |\Lambda|}\sum_{k\in \Lambda^*}[m_{k}\ln m_{k}-(m_{k}+1)\ln(m_{k}+1)]
\label{1.4}
\eea
 for a gas of independent Bose particles having average occupation numbers $m_{k},\; k\in \Lambda^* $.
In the thermodynamic limit,
\bea
s_{0}(m)=\lim_{\Lambda\to\infty}s_{0,\Lambda}(m)=-k_{B}\int \frac{dk}{(2\pi)^{d}}[m_{e}(k)\ln m_{e}(k)-(m_{e}(k)+1)\ln(m_{e}(k)+1)]
\label{1.5}
\eea
only involves the absolutely continuous density $m_{e}(k)$ of the infinite volume mode distribution $m(dk)$ ( for instance it is easily verfied that
a  macroscopically occupied mode, say $n_{k=0}\sim n$, does not contribute to the limit (\ref{1.5});  for the free Bose gas, it is well known that at any finite temperature
the total entropy is entirely due to the non condensed phase).
Adding the chemical potential term, the grand-potential functional of the perturbed mean field model reads
\be
\varphi(m)=u(m)-\mu \la 1,m\ra- Ts_{0}(m)\;.
\label{1.6}
\ee
It is proved in \cite{vdB D L P}  for a class of positive definite interaction kernels $v(k,k')$
that the minimization of $\varphi(m)$ yields indeed the pressure of the perturbed mean field model,
the minimizer $\nu(dk)$ is unique and is of the form  (\ref{0.25}).

We now want to define the entropy density functional $s_{\sQ,\sk \lambda}$ for the statistical ensemble (\ref{0.18}) that has the additional \lq \lq $\lambda\theta_{Q}$ field". For this
we first consider the Legendre transform
\be
\pi^*_{0}(t)=\sup_{ y<0}\(t\,  s -\pi_{0}( s)\)=t  s_{0}(t) -\pi_{0}( y_{0}(t))
\label{1.6a}
\ee
of the function $\pi_{0}( s)$ (\ref{0.30}).
It is given by
\be
\pi^*_{0}(t)=t  s_{0}(t) -\pi_{0}( y_{0}(t))
\label{1.7a}
\ee
where $ y_{0}(t)$ is the unique solution of the equation
\be
t=\pi'( y)= \frac{1}{e^{-\beta y}-1}
\label{1.7}
\ee
and thus
\bea
y_{0}(t)=\beta^{-1}\ln \(\frac{t}{t+1}\),\quad
\pi^*_{0}(t)=-T[-k_{B}(t\ln t-(t+1)\ln(t+1))]\;.
\label{1.8}
\eea
One therefore sees from (\ref{1.5}) that the entropy can be expressed as
\be
 Ts_{0}(m)=-\int \frac{dk}{(2\pi)^{d}} \pi^*_{0}(m_{e}(k))
\label{1.9}
\ee
When $\lambda\neq 0$, this motivates the definition  of an entropy density functional $s_{\sQ,\sk \lambda}(m)$ for the modified ensemble
(depending only on the density of the absolutely continuous part $m_{e}(k)$ of the measure $m(dk)$) by
\bea
Ts_{\sQ,\sk \lambda}(m)=-\int \frac{dk}{(2\pi)^{d}}  \pi^*_{\sQ,\sk \lambda}(m_{e}(k))
\label{1.10}
\eea
where
\be
 \pi^*_{\sQ,\sk \lambda}(t)=\sup_{ y<0}\(t\,  y -\pi_{\sQ,\sk \lambda}( y)\)=ty_{\sQ,\sk \lambda}(t)-\pi_{\sQ,\sk \lambda}(y_{\sQ,\sk \lambda}(t) )
\label{1.11}
\ee
is the Legendre transform of the function (\ref{0.28}) and $y_{\sQ,\sk \lambda}(t)$
is the unique solution in $y$ of the equation
\be
t=\pi'_{\sQ,\sk \lambda}(y).
\label{solution}
\ee
Finally the total functional $\varphi_{\sQ,\sk \lambda}(m)$ (the grand-potential of the modified ensemble)
is taken as
\be
\varphi_{\sQ,\sk \lambda}(m)=u(m)-\mu \la 1,m\ra- Ts_{\sQ,\sk \lambda}(m)
\label{1.12}
\ee
where $u(m)$ and $s_{\sQ,\sk \lambda}(m) $ are defined by (\ref{1.3}) and (\ref{1.10}) respectively.

The corresponding pressure is thus given by
\be
P_{\sQ,\sk \lambda}=-\varphi_{\sQ,\sk \lambda}(\nu_{\sQ,\sk \lambda})
\label{1.13}
\ee
where $\nu_{\sQ,\sk \lambda}(dk)$ is the measure that minimizes $\varphi_{\sQ,\sk \lambda}(m)$.
For $\lambda=0$, $\nu_{\sQ,\sk \lambda=0}(dk)=\nu(dk)$ reduces to the equilibrium mode distribution
(\ref{0.25}) of the perturbed mean field model.
The $\lambda$-derivative of the modified pressure is
\be
\frac{\partial}{\partial \lambda}P_{\sQ,\sk \lambda}\Big|_{\lambda=0}=T\frac{\partial}{\partial \lambda}
s_{\sQ,\sk \lambda}(\nu)\Big|_{\lambda=0}
\label{1.14}
\ee
since derivation with respect to $\nu_{\sQ,\sk \lambda}$ does not contribute at the minimum and only
$s_{\sQ,\sk \lambda}$ explicitly depends on $\lambda$ in $\varphi_{\sQ,\sk \lambda}$.
Therefore we have from (\ref{1.10})
\be
\frac{\partial}{\partial \lambda}P_{\sQ,\sk \lambda}\Big|_{\lambda=0}=\int \frac{dk}{(2\pi)^{d}}\
\frac{\partial}{\partial \lambda} \pi^*_{\sQ,\sk \lambda}(\nu_{e}(k))\Big|_{\lambda=0}
\label{1.15}
\ee
with  $\nu_{e}(k)dk$ the absolutely continuous part of $\nu(dk)$.
From (\ref{1.11}) and (\ref{0.28})
\bea
\frac{\partial}{\partial \lambda} \pi^*_{\sQ,\sk \lambda}(t)\Big|_{\lambda=0}=
\frac{\partial}{\partial \lambda}\pi_{\sQ,\sk \lambda}(y_{0}(t))\Big|_{\lambda=0}=
\pi'_{\sQ,\sk 0}(y_{0}(t))
\eea
and therefore
\be
\frac{\partial}{\partial \lambda}P_{\sQ,\sk \lambda}\Big|_{\lambda=0}=\int \frac{dk}{(2\pi)^{d}}\
\pi'_{\sQ,\sk 0}(y_{0}(\nu_{e}(k)).
\label{1.16}
\ee
From (\ref{1.7})
\bea
\lim_{Q\to\infty}\pi'_{\sQ,\sk 0}(y_{0}(t))=\pi'_{0}(y_{0}(t))=t,
\label{1.17}
\eea
and hence
\be
\lim_{Q\to\infty}\frac{\partial}{\partial \lambda}P_{\sQ,\sk \lambda}\Big|_{\lambda=0}
=\int \frac{dk}{(2\pi)^{d}} \nu_{e}(k).
\label{1.18}
\ee
In view of (\ref{0.22}) this establishes our main result, the identification of the total
excited mode occupation
$\nu_{e}=\int \frac{dk}{(2\pi)^{d}} \nu_{e}(k)$
with the short cycle average number density $\rho_{{\rm short}}$ (\ref{0.17}):
\be
\rho_{{\rm short}}=\nu_{e}.
\label{1.19}
\ee

\section{The large deviation argument}

We shall use the large deviation techniques developed in \cite{vdB D L P}, \cite{E G P} (see also references in \cite{DLP 1})
to justify the variational calculation presented in the previous section.
Here we present the heuristic argument, the technical details are given in the next section.
The  large deviation theory is a generalization of Laplace's method for evaluating the asymptotics
of integrals of the type $\int dp(x)e^{\gamma f(x)}$ with $dp(x)$ a probability measure and $\gamma$
a large parameter. First we have to cast our problem in the form of a Laplace integral.

Let $\Omega_{r,\sLambda}$ to be the set of unordered $r$-tuples
$({\bf q},{\bf k})=\{(q_1,k_1),(q_2,k_2)\dots,(q_r,k_r)\}$ in $\NN\times\Lambda^*$ and let $\Omega_{\sLambda}=\cup_{r\in\NN}\Omega_{r,\sLambda}$.
We introduce a  probability measure on $\Omega_\sLambda$ associated with
the modified ensemble (\ref{0.26}) of the free gas at some negative value $\alpha$ of the chemical potential
\be
\PP^0_{\sQ,\sk\lambda,\sk\sLambda}({\bf q},{\bf k})=\frac{1}{ \Xi^0_{\sQ,\sk\lambda,\sk\sLambda}} \frac{1}{r!}
\prod_{i=1}^r\frac{e^{\beta(\alpha + \lambda\theta_Q(q_i))q_i}}{q_i}
\exp\left[-\beta\sum_{i=1}^r \veps(k_i)q_i\right],\quad \alpha<0
\;.
\label{2.1}
\ee
From (\ref{0.18}) and (\ref{0.24b}) we can now write the total partition function
(normalized with respect to the free one (\ref{0.26}))
as an average with respect to $\PP^0_{\sQ,\sk\lambda,\sk\sLambda}$ of the chemical potential and interaction contributions
\be
\frac{\Xi_{\sQ,\sk\lambda,\sk\sLambda}}{\Xi^0_{\sQ,\sk\lambda,\sk\sLambda}}=
\sum_{{\bf q},{\bf k}\in \Omega_{\sLambda} }\PP^0_{\sQ,\sk\lambda,\sk\sLambda}({\bf q},{\bf k})\exp\(\beta |\Lambda|G(m_{({\bf q},{\bf k})})\)
\label{2.2}
\ee
with
\be
G(m)=(\mu-\alpha) \la 1,m\ra-\half a \la 1,m\ra^2-\half\la m,vm\ra
\label{2.6}
\ee
It is understood in the sequel that quantities pertaining to the free gas (with index $0$)
are taken at an arbitrarily fixed negative value $\alpha$
of the chemical potential (since the latter quantities are not defined when $\alpha$ is positive).
The complete partition $\Xi_{\sQ,\sk\lambda,\sk\sLambda}$ function depends
on the actual value $\mu$, $\mu\neq \alpha$, of the chemical potential. This produces
the term $(\mu-\alpha)\la 1,m\ra$ in (\ref{2.6}).
The $\alpha$ dependence will drop out in the final calculation.

We observe that the function to be averaged in (\ref{2.2}) depends on ${\bf q},{\bf k}$
only through the measure $m_{({\bf q},{\bf k})}$. We can define a measure
$\KK^0_{\sQ,\sk\lambda,\sk\sLambda}(dm)$ on $E$, the space of
positive bounded measures on $\RR^d$, through the relation
\be
\int_{E}F(m)\KK^0_{\sQ,\sk\lambda,\sk\sLambda}(dm)=\sum_{{\bf q},{\bf k}\in \Omega_{\sLambda} }\PP^0_{\sQ,\sk\lambda,\sk\sLambda}({\bf q},{\bf k})
F(m_{({\bf q},{\bf k})})
\label{2.3}
\ee
so that in terms of this measure the expectation (\ref{2.2}) takes the form of an  integral
on the space $E$
\be
\frac{\Xi_{\sQ,\sk\lambda,\sk\sLambda}}{\Xi^0_{\sQ,\sk\lambda,\sk\sLambda}}=\int_E \exp\(\beta|\Lambda| G(m)\)\KK^0_{\sQ,\sk\lambda,\sk\sLambda}(dm)
\label{2.5}
\ee
Then the pressure of the modified ensemble, using (\ref{2.5}), can be expressed in the form
\bea
P_{\sQ,\sk \lambda}&=&\lim_{\Lambda\to \infty}\frac{1}{\beta|\Lambda|} \ln\Xi_{\sQ,\sk\lambda,\sk\sLambda}\nonumber\\
&=&
\lim_{\Lambda\to \infty}\frac{1}{\beta|\Lambda|}\ln\int_E \exp\(\beta|\Lambda| G(m)\)\KK^0_{\sQ,\sk\lambda,\sk\sLambda}(dm)
+ P^{0}_{\sQ,\sk \lambda}
\label{2.7}
\eea
where the last term is the free (modified) pressure (\ref{0.29}).

In a mathematical sense which will be made precise in the next section,
 the measure $\KK^0_{\sQ,\sk\lambda,\sk\sLambda}(dm)$ on $E$
behaves or large $\Lambda$ as
\be
\KK_{\sQ,\sk\lambda,\sk\sLambda}(dm)\approx \exp\(-\beta|\Lambda|I^0_{\sQ,\sk \lambda}(m)\)(dm)
\label{2.8}
\ee
where $I^0_{\sQ,\sk \lambda}(m)$ is called the rate function.
Introducing this asymptotic behaviour in (\ref{2.5}), we see that the partition function
$\Xi_{\sQ,\sk\lambda,\sk\sLambda}$ has the form of a Laplace integral (up to a normalization).
Then by a generalization of Laplace's Theorem (Varadhan's Theorem) \cite{V} we get
\be
P_{\sQ,\sk \lambda}=\sup_{m\in E} \(G(m)-I^0_{\sQ,\sk \lambda}(m)\)+P^{0}_{\sQ,\sk \lambda}.
\label{2.9a}
\ee
We still have to determine the rate function $I^0_{\sQ,\sk \lambda}$.
One can guess it with the help of a well known argument. For a test function $f$ on $\RR^\nu$ consider
\be
C^0_{\sQ,\sk \lambda}(f)=\lim_{\Lambda\to \infty}\frac{1}{\beta|\Lambda|}\ln \int_E \exp\(\beta|\Lambda|\la f, m\ra\)\KK^0_{\sQ,\sk\lambda,\sk\sLambda}(dm)
\label{2.9}
\ee
where $\la f, m\ra=\int f(k)m(dk)$. Applying now Varadhan's Theorem to
(\ref{2.9}) one obtains
\be
C^0_{\sQ,\sk \lambda}(f)=\sup_m \(\la f, m\ra-I^0_{\sQ,\sk \lambda}(m) \),
\label{2.10}
\ee
that is $C^0_{\sQ,\sk \lambda}$ is the Legendre-Fenchel transform of $I^0_{\sQ,\sk \lambda}$.
Inverting this transform one gets
\be
I^0_{\sQ,\sk \lambda}(m)=\sup_f \(\la f, m\ra- C^0_{\sQ,\sk \lambda}(f)\).
\label{2.11}
\ee
We observe that the integral occurring in (\ref{2.9}) has the same form as that in (\ref{2.5})
with $G(m)$ replaced by $\la f, m\ra$. Coming back to expectations
written in terms of the probability $\PP^0_{\sQ,\sk\lambda,\sk\sLambda}({\bf q},{\bf k})$
this amounts to replacing in (\ref{2.2}) the partition function
$\Xi_{\sQ,\sk\lambda,\sk\sLambda}$ by that of the free gas (\ref{0.26})
with shifted energies $\veps(k)-f(k)$
\bea
&&\int_E \exp\(\beta|\Lambda|\la f, m\ra\)\KK^0_{\sQ,\sk\lambda,\sk\sLambda}(dm)
\nonumber\\
&=&\frac{1}{\Xi^0_{\sQ,\sk\lambda,\sk\sLambda}}\sum_{r=1}^{\infty}\frac{1}{r!}\ \sum_{{\mathbf q}\in\NN^r}\ \prod_{i=1}^r\frac{e^{\beta(\alpha+ u\theta_Q(q_i))q_i}}{q_i}
\sum_{{\bf k}\in(\Lambda^*)^r}\exp\left\{-\beta\sum_{i=1}^r \(\veps(k_i)-f(k_i)\)q_i\right\}\;.
\label{2.12}
\eea
By the same calculation that led to (\ref{0.29}) one eventually gets
\be
C^0_{\sQ,\sk \lambda}(f)=\int\frac{dk}{(2\pi)^{d}}\pi_{\sQ,\sk \lambda}(\alpha -\veps(k)+f(k))+P^{0}_{\sQ,\sk \lambda}
\label{2.13}
\ee
To find the inverse Legendre transform of $C^0_{\sQ,\sk \lambda}(f)$ it is useful
to first single out the singular part $m_{c}(dk)$ of $m(dk)$ leading to (see next section)
\be
I^0_{\sQ,\sk \lambda}(m)= -\int (\alpha-\varepsilon(k))m_{c}(dk)+
\sup_f \(\la f, m_{e}\ra- C^0_{\sQ,\sk \lambda}(f)\)+P^{0}_{\sQ,\sk \lambda}
\label{2.14}
\ee
The second term in the r.h.s only involves the density $m_{e}(k)$ of the absolutely
continuous part of $m(dk)$. With the change $g(k)=\alpha-\varepsilon(k)+f(k)$
and in view of (\ref{1.11}) it is equal to
\bea
&& -\int\frac{dk}{(2\pi)^{d}}\ (\alpha-\varepsilon(k))m_{e}(k) + \sup_{g}\int \frac{dk}{(2\pi)^{d}}\ [(g(k)m_{e}(k)
-\pi_{\sQ,\sk \lambda}(g(k))]\nonumber\\
&=&-\int \frac{dk}{(2\pi)^{d}}\ (\alpha-\varepsilon(k))m_{e}(k) +\int\frac{dk}{(2\pi)^{d}}\ \pi^{*}_{\sQ,\sk \lambda}(m_{e}(k))
\label{2.15}
\eea
Finally,
\be
I^0_{\sQ,\sk \lambda}(m)=-\int (\alpha-\varepsilon(k))m(dk) +\int\frac{dk}{(2\pi)^{d}}\ \pi^{*}_{\sQ,\sk \lambda}(m_{e}(k))+P^{0}_{\sQ,\sk \lambda}
\label{2.16}
\ee
and from (\ref{2.9a}) and (\ref{2.6})
\be
P_{\sQ,\sk \lambda}=-\inf_{m\in E}\(\la \veps-\mu ,m\ra +\half a\la 1,m\ra^2+\half\la m,vm\ra+
\int\frac{dk}{(2\pi)^{d}}\ \pi^{*}_{\sQ,\sk \lambda}(m_{e}(k))\)
\label{2.17}
\ee
is precisely the variational problem for the modified ensemble treated in Section 4.

\section{Proofs}
In this section we establish rigorously the identity (\ref{1.16}) which leads to our main result
(\ref{1.19}). This is done in Theorem \ref{Theorem A}.
Recall that for $Q\in \NN $, $ y<0$ and $ \lambda\geq 0$, we have (see \ref{0.28})
\be
\pi_{\sQ,\sk \lambda}(y)=\sum_{q=1}^\sQ \frac{e^{\beta(y+\lambda)q}}{\beta q}
+\sum_{q=\sQ+1}^\infty \frac{e^{\beta y q}}{\beta q}\non
\ee
and
\be
\pi'_{\sQ,\sk \lambda}(y)=\sum_{q=1}^\sQ e^{\beta(y+\lambda)q}+\sum_{q=\sQ+1}^\infty e^{\beta y q}.
\ee
Here $\pi'_{\sQ,\sk \lambda}$ denotes $(\partial\pi_{\sQ,\sk \lambda})/(\partial  y)$.
The function
$ y \mapsto \pi'_{\sQ,\sk \lambda}(y)$ is strictly increasing and its range is $(0,\infty)$. On the other hand
$\pi^*_{\sQ,\sk \lambda}(t)$, the Legendre transform of $\pi_{\sQ,\sk \lambda}$,
is decreasing in $\lambda$ since (see (\ref{1.11}))
\be
\frac{\partial \pi^*_{\sQ,\sk \lambda}}{\partial \lambda}(t)
=-\frac{\partial \pi_{\sQ,\sk \lambda}}{\partial \lambda}( y_{\sQ,\sk \lambda}(t))
=-\sum_{q=1}^\sQ e^{\beta(y_{\sQ,\lambda}(t)+\lambda)q}<0.
\label{part pi*}
\ee
$\pi^*_{\sQ,\sk \lambda}$ is of course convex in $t$
but it is also concave in $\lambda$ as can be seen from the following argument:
\nl
We have
\be
\frac{\partial^2 \pi^*_{\sQ,\sk \lambda}}{\partial \lambda^2}(t)=-\frac{\partial^2 \pi_{\sQ,\sk \lambda}}{\partial \lambda^2}( y_{\sQ,\sk \lambda}(t))
-\frac{\partial \pi'_{\sQ,\sk \lambda}}{\partial \lambda}( y_{\sQ,\sk \lambda}(t))\frac{\partial  y_{\sQ,\sk \lambda}(t)}{\partial \lambda}.
\label{partial 2 lambda1}
\ee
From (\ref{solution}) we get
\be
0=\frac{\partial \pi'_{\sQ,\sk \lambda}}{\partial \lambda}( y_{\sQ,\sk \lambda}(t))
-\pi''_{\sQ,\sk \lambda}( y_{\sQ,\sk \lambda}(t)) \frac{\partial  y_{\sQ,\sk \lambda}(t)}{\partial \lambda}
\ee
which together with (\ref{partial 2 lambda1}) gives
\be
\frac{\partial^2 \pi^*_{\sQ,\sk \lambda}}{\partial \lambda^2}(t)
=- \frac{\ds\frac{\partial^2 \pi_{\sQ,\sk \lambda}}{\partial \lambda^2}( y_{\sQ,\sk \lambda}(t))\pi''_{\sQ,\sk \lambda}
( y_{\sQ,\sk \lambda}(t))
-\(\frac{\partial \pi'_{\sQ,\sk \lambda}}{\partial \lambda}( y_{\sQ,\sk \lambda}(t))\)^2}{\ds\pi''_{\sQ,\sk \lambda}
( y_{\sQ,\sk \lambda}(t))}.
\label{partial 2 lambda2}
\ee
The right hand side of the last identity can be seen to be negative from the explicit expressions:
\bea
\pi''_{\sQ,\sk \lambda}(y)&=&\beta\sum_{q=1}^\sQ qe^{\beta( y+\lambda)q}+\beta\sum_{q=\sQ+1}^\infty q
e^{\beta y q},\non\\
\frac{\partial \pi'_{\sQ,\sk \lambda}}{\partial \lambda}(y)&
=&\ds\frac{\partial^2 \pi_{\sQ,\sk \lambda}}{\partial \lambda^2}(y)
=\beta\sum_{q=1}^\sQ qe^{\beta( y+\lambda)q}.
\eea
From Theorems 2 and 3 in \cite{E G P} we can deduce the next proposition. We note that condition
(1.1) in \cite{E G P} is not necessary; one can replace it with the condition that $\pi'_{\sQ,\sk \lambda}$
is invertible on  $(-\infty, 0)$ (see the proof of Theorem 1(a) on page 681).
\begin{proposition}
If $E$ is equipped withe narrow topology, then the sequence of probability measures
$\KK^0_{\sQ,\sk \lambda,\sk \sLambda}$
on $E$ (defined in (\ref{2.3})) satisfies the large deviation principle
with rate function $I^0_{\sQ,\sk \lambda}$ (defined in (\ref{2.14})) and constants $\beta|\Lambda|$.
\end{proposition}
Explicitly Proposition 1 means that:
\nl
(LD1)\ \ \ $I^0_{\sQ,\sk \lambda}$ is lower semi-continuous;
\nl
(LD2)\ \ \ For each $b<\infty$, the level sets $\{m\in E: I^0_{\sQ,\sk \lambda}(m)\leq b\}$ is compact;
\nl
(LD3)\ \ \ For each closed set $C$
\be
\limsup_{\Lambda\to \infty}(\beta|\Lambda|)^{-1}\ln \KK^0_{\sQ,\sk \lambda,\sk \sLambda}(C)\leq -\inf_{m\in C}I^0_{\sQ,\sk \lambda}(m);\non
\ee
\nl
(LD4)\ \ \ For each open set $O$
\be
\limsup_{\Lambda\to \infty}(\beta|\Lambda|)^{-1}\ln \KK^0_{\sQ,\sk \lambda,\sk \sLambda}(O)\geq -\inf_{m\in O}I^0_{\sQ,\sk \lambda}(m).\non
\ee
\nl
We set $\inf_{m\in \emptyset}I^0_{\sQ,\sk \lambda}(m)=\infty$.
\par
From Lemma 4.1 in \cite{vdB D L P} we know that $m\mapsto G(m)$ (as defined in (\ref{2.6})) is
continuous in the narrow topology.
Then by Varadhan's Theorem \cite{V}, we obtain (\ref{2.17}):
\bea
P_{\sQ,\sk \lambda}(\mu)= -\inf_{m\in E}{\cal E}_{\sQ,\sk \lambda}(m)
\eea
where the functional ${\cal E}_{\sQ,\sk \lambda}:E\to \RR$ is defined by
\be
{\cal E}_{\sQ,\sk \lambda}(m)=\la \veps-\mu ,m\ra +\half a\|m\|^2+\half\la m,vm\ra+
\int\frac{dk}{(2\pi)^{d}}\ \pi^{*}_{\sQ,\sk \lambda}(m_{e}(k)).
\ee
We have the following results:
\ben
\item
${\cal E}_{\sQ,\sk \lambda}(0)=0$.
\item
${\cal E}_{\sQ,\sk \lambda}$ is convex in $m$ and concave, differentiable and decreasing in $\lambda$.
The concavity of ${\cal E}_{\sQ,\sk \lambda}$ in $\lambda$ follows from the
concavity of $\pi^*_{\sQ,\sk \lambda}$ in $\lambda$. The latter together with (\ref{part pi*}) gives
\be
\frac{\partial{\cal E}_{\sQ,\sk \lambda}}{\partial \lambda}(m)
=-\int \frac{dk}{(2\pi)^{d}}\
\frac{\partial \pi_{\sQ,\sk \lambda}}{\partial \lambda}( y_{\sQ,\sk \lambda}(m_{e}(k)))dk
\ee
which implies that
\be
-\|m\| \leq \frac{\partial{\cal E}_{\sQ,\sk \lambda}}{\partial \lambda}(m)\leq 0.
\ee
since
\be
0\leq \frac{\partial \pi_{\sQ,\sk \lambda}}{\partial \lambda}( y_{\sQ,\sk \lambda}(t))
\leq \pi'_{\sQ,\sk \lambda}( y_{\sQ,\sk \lambda}(t))=t.
\ee
From the continuity of $G_{\sQ,\sk \lambda}$ and the lower semi-continuity of $I^0_{\sQ,\sk \lambda}$
it follows that ${\cal E}_{\sQ,\sk \lambda}(m)$ is lower semi-continuous in $m$.
\item
Suppose that $v$ is symmetric and is continuously differentiable with bounded derivatives. Then
${\cal E}_{\sQ,\sk \lambda}$ has a unique minimizer $\nu_{\sQ,\sk \lambda}$ whose singular part is concentrated at $0$.
The minimizer satisfies the Euler-Lagrange equations:
\bea
\veps(0)-\mu+a\|\nu_{\sQ,\sk \lambda}\|+(v\nu_{\sQ,\sk \lambda})(0)&=&0,\\
\veps(k)-\mu+a\|\nu_{\sQ,\sk \lambda}\|+(v\nu_{\sQ,\sk \lambda})(k)&
=& -y_{\sQ,\sk \lambda}(\nu_{\sQ,\sk \lambda, \sk e}(k)),\ \ \ \ {\hbox{\rm Lebesgue-a.e.}},
\label{density1}
\eea
where $\nu_{\sQ,\sk \lambda, \sk e}$ is the density of the absolutely continuous part
of $\nu_{\sQ,\sk \lambda}$.
This is proved by the same arguments as in Theorem 6
in \cite{vdB D L P}. Equation (\ref{density1}) implies that
\be
\nu_{\sQ,\sk \lambda, \sk e}(k)=\pi'_{\sQ,\sk \lambda}( -\left\{\veps(k)-\mu+a\|\nu_{\sQ,\sk \lambda}\|
+(v\nu_{\sQ,\sk \lambda})(k)\right \}).
\label{density2}
\ee
\een
We shall prove the following theorem:
\begin{theorem}\label{Theorem A}
\be
\frac{\partial P_{\sQ,\sk \lambda}}{\partial \lambda}\Big|_{\lambda=0}
=\int \hskip -0.1cm \frac{dk}{(2\pi)^{d}}\
\sum_{q=1}^\sQ \exp \{\beta (y_{0}(\nu_{e}(k))q\}=\int \frac{dk}{(2\pi)^{d}}\
\pi'_{\sQ,\sk 0}(y_{0}(\nu_{e}(k)).
\label{ThA}
\ee
\end{theorem}
\begin{lemma}
For every sequence $\{\lambda_n\}$ of positive numbers tending to zero, $\nu_{\sQ,\sk \lambda_n}$
converges to $\nu$ in the narrow topology.
Moreover the sequence $\nu_{\sQ,\sk \lambda_n}$ is tight and
$\nu_{\sQ,\sk \lambda_n,\sk e}(k)$ converges pointwise to $\nu_e(k)$.
\end{lemma}
\begin{proof}
First we prove that the set of minimizers $\{\nu_{\sQ,\sk \lambda}\ |\ 0\leq \lambda \leq 1\}$
lies inside a compact set.
To do this we follow the proof of Lemma 5.1 in
\cite{vdB D L P} using the additional fact that for fixed $m$, ${\cal E}_{\sQ,\sk \lambda}(m)$ is
decreasing in $\lambda$.
Since ${\cal E}_{\sQ,\sk \lambda}(0)=0$, ${\cal E}_{\sQ,\sk \lambda}(\nu_{\sQ,\sk \lambda})\leq 0$
and thus
${\cal E}_{\sQ,1}(\nu_{\sQ,\sk \lambda})\leq {\cal E}_{\sQ,\sk \lambda}(\nu_{\sQ,\sk \lambda})\leq 0$
for $0\leq \lambda\leq 1$ .
Therefore
\bea
I^\alpha_{\sQ,\sk 1}(\nu_{\sQ,\sk \lambda})&=& {\cal E}_{\sQ,1}(\nu_{\sQ,\sk \lambda})
+G(\nu_{\sQ,\sk \lambda})+P^0_{\sQ,\sk1}\non\\
&\leq & G(\nu_{\sQ,\sk \lambda})\leq (\mu-\alpha)\|\nu_{\sQ,\sk \lambda}\|
- \half a \|\nu_{\sQ,\sk \lambda}\|^2
+P^0_{\sQ,\sk 1}\non\\
&= & - \frac{1}{2a}\{ a \|\nu_{\sQ,\sk \lambda}\| -(\mu-\alpha)\}^2+ \frac{1}{2a}(\mu-\alpha)^2\non\\
&\leq &\frac{1}{2a}(\mu-\alpha)^2+P^0_{\sQ,\sk 1}
\eea
which implies that $\nu_{\sQ,\sk \lambda}$ lies in a level set of $I^\alpha_{\sQ,\sk 1}$
which is compact by LD2.
\par
Let $\{\lambda_n\}$ be a sequence of positive numbers tending to zero and let $\nu^*$ be a limit point of
$\{\nu_{\sQ,\sk \lambda_n}\}$.
Then there is a subsequence $ \lambda_{n_l}$ such $\nu_{\sQ,\sk  \lambda_{n_l}}$ converges to $\nu^*$.
For $m\in E$ we have
\be
{\cal E}_{\sQ,\sk  \lambda_{n_l}}(\nu_{\sQ,\sk  \lambda_{n_l}})\leq {\cal E}_{\sQ,\sk  \lambda_{n_l}}(m)\leq {\cal E}_{\sQ,\sk  0}(m)
\ee
and therefore
\be
\liminf_{l\to\infty}{\cal E}_{\sQ,\sk  \lambda_{n_l}}(\nu_{\sQ,\sk  \lambda_{n_l}})\leq {\cal E}_{\sQ,\sk  0}(m).
\ee
Now
\bea
{\cal E}_{\sQ,\sk  \lambda_{n_l}}(\nu_{\sQ,\sk  \lambda_{n_l}})&
=&{\cal E}_{\sQ,\sk  0}(\nu_{\sQ,\sk  \lambda_{n_l}})
-\({\cal E}_{\sQ,\sk  0}(\nu_{\sQ,\sk  \lambda_{n_l}})
-{\cal E}_{\sQ,\sk  \lambda_{n_l}}(\nu_{\sQ,\sk  \lambda_{n_l}})\)\non\\
&\geq &{\cal E}_{\sQ,\sk  0}(\nu_{\sQ,\sk  \lambda_{n_l}})- \lambda_{n_l}\|\nu_{\sQ,\sk  \lambda_{n_l}}\|.
\eea
and thus since ${\cal E}_{\sQ,\sk  0}$ is lower semi-continuous
$\liminf_{l\to\infty}{\cal E}_{\sQ,\sk  \lambda_{n_l}}(\nu_{\sQ,\sk  \lambda_{n_l}})\geq {\cal E}_{\sQ,\sk  0}(\nu^*)$
and so ${\cal E}_{\sQ,\sk  0}(\nu^*)\leq {\cal E}_{\sQ,\sk  0}(m)$ for all $m\in E$.
This implies that $\nu^*=\nu_{\sQ,\sk  0}=\nu$ and therefore $\{\nu_{\sQ,\sk \lambda_n}\}$
has only one limit point $\nu$ and thus  $\nu_{\sQ,\sk \lambda_n}$ converges to $\nu$.
From (\ref{density2}) using $\pi'_{\sQ,\sk \lambda}(y)\leq \pi'_{\sQ,\sk  0}(y+\lambda)$
with, $v_0=\sup|v(k,k')|$, we have that
\bea
\nu_{\sQ,\sk \lambda_n,\sk e}(k)
&\leq &\pi'_{\sQ,\sk  0}( -\left\{\veps(k)-\mu+a\|\nu_{\sQ,\sk \lambda_n}\|
+(v\nu_{\sQ,\sk \lambda_n})(k)-1\right \})\non\\
&\leq&\pi'_{\sQ,\sk  0}( -\left\{\veps(k)-\mu-v_0\|\nu_{\sQ,\sk \lambda_n}\|\right \}-1)\non\\
&\leq&\pi'_{\sQ,\sk  0}( -\left\{\veps(k)-\mu-v_0K-1\right \})
\eea
for some $K$ and all $n$. For large $\|k\|$, $\veps(k)-\mu-v_0K-1>\veps(k)/2$ and therefore
\be
\nu_{\sQ,\sk \lambda_n,\sk e}(k)\leq \frac{1}{e^{\beta \veps(k)/2}-1}.
\ee
Again from (\ref{density2}) that $\nu_{\sQ,\sk \lambda_n,\sk e}(k)$ converges pointwise to
\be
\pi'_{\sQ,\sk  0}( -\left\{\veps(k)-\mu+a\|\nu\|+(v\nu)(k)\right \})
\ee
since $(v\nu_{\sQ,\sk \lambda_n})(k)\to (v\nu)(k)$ for each $k$.
\end{proof}
{\bf Proof of Theorem \ref{Theorem A}}
\par
Recall that $P_{\sQ,\sk \lambda}(\mu)=-{\cal E}_{\sQ,\sk \lambda}(\nu_{\sQ,\sk \lambda})$. Therefore
\bea
P_{\sQ,\sk \lambda_n}(\mu)-P_{\sQ,\sk  0}(\mu)
&=&{\cal E}_{\sQ,\sk  0}(\nu_{\sQ,\sk  0})-{\cal E}_{\sQ,\sk \lambda_n}(\nu_{\sQ,\sk \lambda_n})\non\\
&=&({\cal E}_{\sQ,\sk  0}(\nu_{\sQ,\sk  0})-{\cal E}_{\sQ,\sk \lambda_n}(\nu_{\sQ,\sk  0}))
+({\cal E}_{\sQ,\sk \lambda_n}(\nu_{\sQ,\sk  0})-{\cal E}_{\sQ,\sk \lambda_n}(\nu_{\sQ,\sk \lambda_n}))\non\\
&\geq &{\cal E}_{\sQ,\sk  0}(\nu_{\sQ,\sk  0})-{\cal E}_{\sQ,\sk \lambda_n}(\nu_{\sQ,\sk  0})
\geq -\lambda_n \frac{\partial {\cal E}_{\sQ,\sk \lambda}(\nu_{\sQ,\sk  0})}{\partial \lambda}
\Big |_{\lambda=0}.
\eea
In the last inequality we have used the fact that ${\cal E}_{\sQ,\sk \lambda}$ is concave in $\lambda$.
Therefore
\bea
\liminf_{n \to \infty}\frac{P_{\sQ,\sk \lambda_n}(\mu)-P_{\sQ,\sk  0}(\mu)}{\lambda_n}
&\geq&\int \hskip -0.1cm \frac{dk}{(2\pi)^{d}}\
\frac{\partial \pi_{\sQ,\sk \lambda}}{\partial \lambda}( y_{\sQ,\sk \lambda}(\nu_{\sQ,\sk 0,\sk e}(k))
\Big |_{\lambda=0}\non\\
&=&\int \hskip -0.1cm \frac{dk}{(2\pi)^{d}}\ \sum_{q=1}^\sQ
\exp \{\beta y_{\sQ,\sk  0}(\nu_{\sQ,\sk 0,\sk e}(k))q\}\non\\
&=&\int \hskip -0.1cm \frac{dk}{(2\pi)^{d}}\ \sum_{q=1}^\sQ
\exp \{\beta y_0(\nu_e(k))q\}.
\eea
On the other hand
\bea
P_{\sQ,\sk \lambda_n}(\mu)-P_{\sQ,\sk  0}(\mu)&
=&{\cal E}_{\sQ,\sk  0}(\nu_{\sQ,\sk  0})-{\cal E}_{\sQ,\sk \lambda_n}(\nu_{\sQ,\sk \lambda_n})\non\\
&=&({\cal E}_{\sQ,\sk  0}(\nu_{\sQ,\sk  0})-{\cal E}_{\sQ,\sk  0}(\nu_{\sQ,\sk \lambda_n}))
+({\cal E}_{\sQ,\sk  0}(\nu_{\sQ,\sk \lambda_n})-{\cal E}_{\sQ,\sk \lambda_n}(\nu_{\sQ,\sk \lambda_n}))\non\\
&\leq &{\cal E}_{\sQ,\sk  0}(\nu_{\sQ,\sk \lambda_n})
-{\cal E}_{\sQ,\sk \lambda_n}(\nu_{\sQ,\sk \lambda_n})\leq -\lambda_n
\frac{\partial {\cal E}_{\sQ,\sk \lambda}(\nu_{\sQ,\sk \lambda_n})}{\partial \lambda}
\Big |_{\lambda=\lambda_n}.
\eea
Thus
\bea
\limsup_{n \to \infty}\frac{P_{\sQ,\sk \lambda_n}(\mu)-P_{\sQ,\sk  0}(\mu)}{\lambda_n}
\leq\int \hskip -0.1cm \frac{dk}{(2\pi)^{d}}\ \sum_{q=1}^\sQ
\exp \{\beta\(\lambda_n+ y_{\sQ,\sk \lambda_n}(\nu_{\sQ,\sk \lambda_n,\sk e}(k))\)q\}.
\label{upper bnd}
\eea
To show that the righthand side of (\ref{upper bnd}) converges to the righthand side of (\ref{ThA})
we note that
\be
\sum_{q=1}^\sQ \exp \{\beta\(\lambda_n+ y_{\sQ,\sk \lambda_n}(\nu_{\sQ,\sk \lambda_n,\sk e}(k))\)q\}
\leq \pi'_{\sQ,\sk \lambda_n}( y_{\sQ,\sk \lambda_n}(\nu_{\sQ,\sk \lambda_n,\sk e}(k)))
=\nu_{\sQ,\sk \lambda_n,\sk e}(k).
\ee
Therefore by using the tightness of the sequence $\nu_{\sQ,\sk \lambda_n}$ we can make both the integrals
\be
\int_{\|k\|^2>R}\hskip -0.1cm dk \sum_{q=1}^\sQ
\exp \{\beta\(\lambda_n+ y_{\sQ,\sk \lambda_n}(\nu_{\sQ,\sk \lambda_n,\sk e}(k))\)q\},
\ \ \ \ \ \int_{\|k\|^2>R}\hskip -0.1cm dk \sum_{q=1}^\sQ \exp \{\beta y_{\sQ,\sk  0}(\nu_{\sQ,\sk 0,\sk e}(k))q\}
\ee
arbitrarily small by taking $R$ large enough. On $\|k\|^2\leq R$ we use the bound
\be
\sum_{q=1}^\sQ \exp \{\beta\(\lambda_n+ y_{\sQ,\sk \lambda_n}(\nu_{\sQ,\sk \lambda_n,\sk e}(k))\)q\}\leq
\sum_{q=1}^\sQ \exp \{\beta q\}.
\ee
for $\lambda_n\leq 1$ and the Lebesgue Dominated Convergence Theorem.
\par
\qed

\section{Concluding remarks}

The most general way to define the condensate density $\rho_{c}$ for an interacting Bose
system is by means of the concept of off-diagonal long range order
\be
\rho_{c}=\lim_{|x-x'|\to\infty}\lim_{|\sLambda |\to\infty}\rho_{\sLambda}(x,x')
\label{3.1}
\ee
where  $\rho_{\sLambda}(x,x')$ are the configurational matrix elements of the
one-body reduced density matrix.
For the perturbed mean field model the infinite volume reduced density matrix $\rho(x,x')$
is simply the Fourier transform of the equilibrium distribution $\nu(dk)$ (\ref{0.25})
so that
\be
\lim_{|x-x'|\to\infty}\rho(x,x')=\lim_{|x-x'|\to\infty}\int \nu(dk)e^{ik\cdot (x-x')}=\nu_{c}
\label{3.2}
\ee
since the Fourier transform of the smooth part of the mode distribution does not contribute
asymptotically.
We therefore see that the three possible definitions of condensation, namely particles in long cycles $\rho_{{\rm long}}$, macroscopic occupation of the $k=0$ mode $\nu_{c}$
and off-diagonal long range order $\rho_{c}$ coincide in this model.
It may be conjectured that the relation $\rho_{{\rm long}}=\rho_{c}$ holds under rather general circumstances
(e.g. for models with local pair interactions). It is therefore of interest to examine the relation of $\rho_{\sLambda}(x,x')$
with the cycle statistics. For this we first consider the average of an arbitrary one-particle observable
of the form $A(n)=\sum_{i=1}^{n}A_{i}$ with $A_{i}=A$ a single particle operator
acting on the one-particle Hilbert space ${\cal H}_\sLambda$.
Its grand-canonical average reads
\be
\langle A\rangle_{\sLambda}=\frac{1}{\Xi^\mu_\sLambda}\sum_{n=1}^{\infty}e^{\beta \mu n}
\tr_{{\cal H}^{\otimes n}_\sLambda}\left[A(n)e^{-\beta H_\sLambda(n)}\right]
\label{a1}
\ee
where $H_\sLambda(n)$ is the $n$-particle Hamiltonian. We can write as in (\ref{0.2})
\bea
&&\tr_{{\cal H}^n_{\sLambda,{\tiny \rm symm}}}\left[A(n)e^{-\beta H_\sLambda(n)}\right]=
\frac{1}{n!}\sum_{\pi\in S_n}f^{A}_{\sLambda}(\pi)\quad\quad{\rm with}\nonumber\\
&&f_{\sLambda}^{A}(\pi)=\tr_{{\cal H}^{\otimes n}_\sLambda}\left[U(\pi)A(n)e^{-\beta H_\sLambda(n)}\right]\;.
\label{a2}
\eea
For the same reasons as in (\ref{0.5}) $f_{\sLambda}^{A}(\pi)=f_{\sLambda}^{A}({\bf q})$ only depends
on the cycle decomposition of $\pi$ and all the steps leading to (\ref{0.8}) are the same:
\be
 \langle A\rangle_{\sLambda}=\frac{1}{\Xi^\mu_\sLambda}
\sum_{r=1}^{\infty}\frac{1}{r!}\ \sum_{{\mathbf q}\in\NN^r}\ \prod_{i=1}^r\frac{e^{\beta\mu q_i}}{q_i} f_\sLambda^{A}(q_1,q_2,\dots,q_r)\;.
\label{a3}
\ee
With (\ref{0.5b}) one can write the function $f_\sLambda^{A}(q_1,q_2,\dots,q_r)$ more explicitly as
\be
f_\sLambda^{A}(q_1,q_2,\dots,q_r)=\tr_{{\cal H}^{\otimes q_{1}}_\sLambda\otimes\cdots\otimes{\cal H}^{\otimes q_{r}}_\sLambda}
\left[U_{q_{1}}U_{q_{2}}\cdots U_{q_{r}}\sum_{j=1}^{r}A(q_{j})e^{-\beta H_\sLambda(q_1,q_2,\dots,q_r)}\right]
\label{a41}
\ee
where the notation $H_\sLambda(n)=H_\sLambda(q_1,q_2,\dots,q_r)$ indicates
that among the $n$ interacting particles $q_{j}$ of them belong to the cycle $j,\;j=1,\ldots,r$.
It is a symmetric function of $q_1,q_2,\dots,q_r$ so that when inserted in (\ref{a3}) and relabelling
the dummy summation indices, it contributes as
\bea
&&r \times\tr_{{\cal H}^{\otimes q_{1}}_\sLambda\otimes\cdots\otimes{\cal H}^{\otimes q_{r} }}\left[
U_{q_{1}}U_{q_{2}}\cdots U_{q_{r}}A(q_{1})e^{-\beta H_\sLambda(q_1,q_2,\dots,q_r)}\right]
=\nonumber\\  && r\times\tr_{
{\cal H}^{\otimes q_{1}}_\sLambda}\left[U_{q_{1}}A(q_{1})\tr_{
{\cal H}^{\otimes q_{2}}_\sLambda\otimes\cdots\otimes{\cal H}^{\otimes q_{r}}_\sLambda
}\left(
U_{q_{2}}\cdots U_{q_{r}}e^{-\beta H_\sLambda(q_1,q_2,\dots,q_r)}\)\right]
\label{a4}
\eea
Hence we obtain finally (changing $r\to r+1$ in the $r$-sum)
\be
 \langle A\rangle_{\sLambda}=\sum_{q=1}^{\infty}\tr_{{\cal H}^{\otimes q}_\sLambda}
\left[A(q)U_{q}R_{\sLambda}(q)\right]
\label{a5}
\ee
where we have introduced the cycle reduced density matrix $R_{\sLambda}(q)$ as an operator acting
in the space  ${\cal H}^{\otimes q}_\sLambda$ of the the particles belonging to
a $q$-cycle:
\be
R_{\sLambda}(q)=\frac{e^{\beta\mu q}}{q}\;\frac{1}{\Xi^\mu_\sLambda}
\sum_{r=1}^{\infty}\frac{1}{r!}\hskip -0.1cm  \sum_{{q_1,q_2,\dots,q_r}\in\NN^r}\hskip -0.02cm \prod_{i=1}^r\frac{e^{\beta\mu q_i}}{q_i}
\tr_{{\cal H}^{\otimes q_{1}}_\sLambda\otimes\cdots\otimes{\cal H}^{\otimes q_{r}}_\sLambda}
\left[
U_{q_{1}}U_{q_{2}}\cdots U_{q_{r}}e^{-\beta H_\sLambda(q,q_1,q_2,\dots,q_r)}\right]
\label{a6}
\ee
The reduced density matrix is obtained from (\ref{a5}) by specifying
$A(q)=\sum_{i=1} ^{q}P_{i}(x,x')$ with $P_{i}(x,x')$ acting as $P(x,x')=| x\ra\la x'|$ on the one-particle space
\be
\rho_{\sLambda}(x,x')=
\sum_{q=1}^{\infty}\tr_{{\cal H}^{\otimes q}_\sLambda}
\left[\sum_{i=1} ^{q}P_{i}(x,x')U_{q}R_{\sLambda}(q)\right]\;.
\label{a7}
\ee
We want to show that, as for $\rho_{{\rm long}}$, off-diagonal long range order is carried by infinitely long cycles.
For this consider the case of a
positive pair potential and let $H_{\sLambda}^{0}(q)=\sum_{i}^{q}||p_{i}||^{2}/2M$ be the kinetic energy of $q$ particles. Then by dropping the interactions of the particles
in the $q$-cycle with those belonging to other cycles $q_{j}, j=1,\ldots,r$, one has the inequalities
\bea
&&H_{\sLambda}(q,q_{1},\ldots,q_{r})\leq H_{\sLambda}^{0}(q)
+H_{\sLambda}(q_{1},\ldots,q_{r})\nonumber\\
&&
\tr_{{\cal H}^{\otimes q_{1}}_\sLambda\otimes\cdots\otimes{\cal H}^{\otimes q_{r}}_\sLambda}
\left[
U_{q_{1}}U_{q_{2}}\cdots U_{q_{r}}e^{-\beta H_\sLambda(q,q_1,q_2,\dots,q_r)}\right]
\leq\nonumber\\&&\quad\quad\quad
e^{-\beta H_{\sLambda}^{0}(q)}\tr_{{\cal H}^{\otimes q_{1}}_\sLambda\otimes\cdots\otimes{\cal H}^{\otimes q_{r}}_\sLambda}
\left[
U_{q_{1}}U_{q_{2}}\cdots U_{q_{r}}e^{-\beta H_\sLambda(q_1,q_2,\dots,q_r)}\right]\;.
\label{a8}
\eea
The above traces involve off-diagonal configurational matrix elements because of the presence of the permutation operators $U_{q_{j}}$
but the inequality is easily seen to hold using the Feynman-Kac path integral representation of these traces.
When inserted in (\ref{a6}) this shows that each term of the sum (\ref{a7}) is bounded by
\bea
&& \frac{e^{\beta \mu q}}{q}
\tr_{{\cal H}^{\otimes q}_\sLambda}
\left[\sum_{i=1} ^{q}P_{i}(x,x')U_{q}e^{-\beta H_{\sLambda}^{0}(q)}\right]\nonumber\\
&= &e^{\beta \mu q}\la x | e^{-\beta q H_{\sLambda}^{0}}|x'\ra
\leq  e^{\beta \mu q}\(\frac{M}{2\pi \hbar^{2}\beta q}\)^{d/2}\exp\(
-\frac{M||x-x'||^{2}}{2\hbar^2\beta q}\)\;.
\label{a9}
\eea
For the last upper bound we have used that Dirichlet boundary conditions increase the kinetic
energy and written the explicit
form of the kernel of $e^{-\beta H^{0}(q)}$ in infinite space.
Therefore each term of the sum (\ref{a7}) tends to zero uniformly with respect to $\Lambda$
as $|x-x'|\to\infty$ and we conclude that only the tail of the series (\ref{a7}) will
contribute to off-diagonal long range order, namely
\be
\rho_{c}=\lim_{Q\to\infty}\lim_{|x-x'|\to\infty} \lim_{|\sLambda|\to \infty}
\sum_{q=Q}^{\infty}\tr_{{\cal H}^{\otimes q}_\sLambda}
\left[\sum_{i=1} ^{q}P_{i}(x,x')U_{q}R_{\sLambda}(q)\right]\;.
\label{a10}
\ee
At the moment we do not have a proof of the equality of the expressions (\ref{a10}) and (\ref{0.16})
for a Bose gas with pair potential interactions.

In numerical calculations of the path integral representation of the Bose gas \cite{Ceperley},
it is more convenient to use periodic
boundary conditions, i. e. $\Lambda$ is a $d$-dimensional torus. Cycles with $q$
particles correspond to Brownian paths on the torus with typical length $\sim \sqrt{ q}$.
To such a path one can associate a winding number, the number of times that the path wraps around the torus.
Then the typical length of a winding path is at least of the order $\sqrt{ q}\sim |\Lambda|^{1/d}$,
the linear dimension of $\Lambda$ and hence the number of paths such that $q\geq
|\Lambda|^{2/d}$ is larger than the number of winding paths.
For a particle configuration , let $N_{w}$ be the total number of particles found in winding paths.
Since $q\rho_{\sLambda}(q)$ is the average
number density of particles in $q$-cycles, one has for $\Lambda$ sufficiently large
\be
\sum_{q=Q+1}^{\infty}q\rho_\sLambda(q)\geq
\sum_{q\geq |\Lambda|^{2/d}}^{\infty}q\rho_\sLambda(q)\geq \frac{\la N_{w}\ra}{|\Lambda|}\;.
\label{a11}
\ee
If the average number density of particles in winding paths $\la N_{w}\ra/|\Lambda|$
in is different from zero in the infinite volume limit, we conclude that $\rho_{{\rm long}}$ is non zero
and thus Bose condensation occurs (provided that $\rho_{{\rm long}}\neq 0$ is equivalent for
the other possible definitions of the condensate density).
This criterion can be compared with that for superfluidity
involving the fluctuations density of the winding number \cite{Ceperley}.
Thus the statistics of winding paths will make it possible to estimate both the superfluid
density and the Bose condensate.
\par
{\bf Acknowledgements:} The authors TCD and JVP wish to
thank the Institut de Physique Th\'eorique, Ecole Polytechnique F\'ed\'erale de
Lausanne where this work was initiated, for its
hospitality and financial support. JVP  also
thanks University College Dublin for the award of a President's
Research Fellowship. PAM thanks DIAS for kind hospitality and support,
as well as a stimulating discussion with F. Lalo\"e.

%


\begin{thebibliography}{xx}

\bibitem{Fey 1953 and stat.mech.} R. P. Feynman, Atomic theory of the $\lambda$ transition in Helium,
{\it Phys. Rev.}, {\bf 91}, 1291, 1953,
and R. P. Feynman, Statistical Mechanics, Chap. 11, {\it Benjamin}, 1974

\bibitem{Pen.Onsa.}  O. Penrose, L. Onsager, Bose-Einstein condensation and Liquid Helium,
{\it Phys. Rev.}, {\bf 104}, 576, 1956

\bibitem{Chandler} D. Chandler, P. G. Wolynes, Exploiting the isomorphism between quantum theory
and classical statistical mechanics of polyatomic fluids, {\it J. Chem. Phys.}, {\bf 74}, 4078, 1981

\bibitem{Sear-Cuesta} R. P. Sear, J. A. Cuesta, What do emulsification failure and Bose-Einstein
condensation have in common ?, {\it  Europhys. Lett.}, {\bf 55}, 451, 2001

\bibitem{Schakel} A. M. J. Schakel, Percolation, Bose-Einstein condensation and string proliferation,
{\it Phys. Rev. E\,}, {\bf 63}, 126115, 2001

\bibitem{Suto} A. S\"ut\"o, Percolation transition in the Bose gas,
{\it J. Phys. A: Math. Gen.},{\bf 26}, 4689, 1993
and
Percolation transition in the Bose gas: II
{\it J. Phys. A: Math. Gen.}, {\bf 35}, 6995, 2002

\bibitem{vdB D L P} M. van den Berg , T.C. Dorlas , J.T. Lewis  and  J.V. Pul\'e ,
A perturbed mean field model of an interacting boson gas and the Large Deviation Principle,
\textit{Commun. Math. Phys.}, \textbf{127} 4, 1990

\bibitem{Ginibre} J. Ginibre,
\lq\lq Some applications of functional integration in statistical mechanics",
{\it Statistical Mechanics and Quantum Field Theory}, C.
DeWitt and R. Stora, eds, Les Houches, Gordon and Breach, 327, 1971

\bibitem{Cornu} F. Cornu, Correlations in quantum plasmas, {\it Phys. Rev. E\,}, {\bf 53}, 4562, 1996

\bibitem{Ceperley} D. M. Ceperley, Path integrals in the theory of condensed Helium,
{\it Rev. Mod. Phys.}, {\bf 67}, 279, 1995

\bibitem{Martin}  Ph. A. Martin, Quantum Mayer graphs: application to Bose and Coulomb gases,
\textit{Acta Phys. Pol. B}, \textbf{34}, 3629, 2003

\bibitem{DLP 1}  T.C. Dorlas, J.T. Lewis  and J.V Pul\'e, Condensation in Some Perturbed
Meanfield Models of a Bose Gas, {\it  Helv. Phys. Acta}, 1200, 1991

\bibitem{DLP 2}  T.C. Dorlas, T. C. Lewis  and J.V Pul\'e, Condensation in a variational problem
on the space of measures, {\it Arch. Rational Mech. Anal.}, {\bf 118}, 245, 1992

\bibitem{Griffiths} R. B. Griffiths, Peierls Proof of Spontaneous Magnetization in a
Two-Dimensional Ising Ferromagnet, {\it Phys. Rev.}, {\bf 136}, A437, 1964

\bibitem{Landau} L. Landau and E. Lifshitz, Statistical Mechanics, {\it Pergamon Press}, 1976

\bibitem{E G P} R.S. Ellis , J. Gough  and J.V. Pul\'e , The Large Deviation Principle for measures with random weights,
\textit{Rev. Math. Phys.}, \textbf{5}, 659, 1993

\bibitem{V} S.R.S. Varadhan , Asymptotic probabilities and differential equations,
\textit{Commun. Pure Appl. Math.}, \textbf{19}, 261, 1989

\end{thebibliography}
\end{document}